\documentstyle[prl,epsf,aps]{revtex}

\begin{document}
%\draft
%\preprint{}
\wideabs{
\title{
Spectral Flow and the Dynamics
of Dislocations\\ in Charge Density Waves
}
\author{Masahiko HAYASHI}
\address{
Department of Applied Physics/DIMES, 
Delft University of Technology,\\
Lorentzweg 1, 2628 CJ Delft, The Netherlands
}
\date{August 5, 1997: Submitted to PRL}
\maketitle
\begin{abstract}
The spectral flow in the cores of 
moving dislocations is found to play an 
important role in the dynamics and transport 
of charge density wave (CDW) 
by significantly modifying the hydrodynamic effective action 
of the condensate. 
The analogy of spectral flow in CDW with the 
baryogenesis in the early universe is pointed out. 
\end{abstract}
\pacs{71.45.Lr, 61.72.Lk, 98.80.Cq}
}

\narrowtext

\begin{figure}
\begin{center}
{\parbox{6cm}
{\epsfysize=6cm \epsfbox{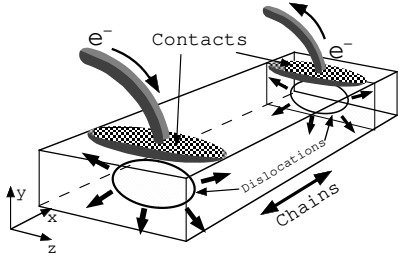}}}
\end{center}
\caption{
A schematic view of the nonlinear conductivity 
measurement. Wave fronts are indicated by shades. 
Dislocation loops and their motions are shown. 
}
\label{pict}
\end{figure}

The dynamics of the dislocation lines in charge density 
waves (CDW's) \cite{review} is one of the key 
processes which govern the sliding (nonlinear) 
conductance of the incommensurate CDW with 
three dimensional (3D) order. 
When current contacts are attached to the side of the chains, 
a sliding CDW implies that wave fronts are added or removed at the 
contacts by the nucleation or 
annihilation of dislocation loops as shown in Fig. \ref{pict}. 
The narrow band noise \cite{Gorkov,Ong-Maki} and 
the nonlinear current-voltage characteristics 
\cite{Ramakurishuna,Maher,Duan,Maki,ZaitzevZotov} have been attributed 
to this phenomena. 
Although this idea explains most qualitative 
features of CDW very well, 
the dynamics of the dislocations is not yet well understood. 
Especially, the gauge invariance or charge conservation condition of 
CDW has not been investigated seriously. 

In this Letter, we first show that 
the gauge invariance of the conventional 
hydrodynamic effective action of the condensate is 
violated in the presence of moving dislocations. 
The origin of this is shown to be 
the neglect of the \lq\lq spectral flow\rq\rq\  
in the cores of the moving dislocations. 
By explicitly taking the effects of 
the spectral flow into account, 
we derive a gauge invariant action 
which correctly describes the dislocation dynamics. 
We find two correction terms arising from the 
spectral flow, namely, 
the correction to the interaction of the 
condensate with the electromagnetic 
field and to the condensation energy. 
The nucleation process of the dislocation loop 
is analyzed based on the obtained action 
and the nonlinear conduction of the CDW is discussed 
in the light of our findings. 
Finally we discuss a remarkable similarity of the 
dislocations in CDW with the so-called cosmic strings in the early 
universe, suggesting the intriguing possibility of 
the CDW system as a theoretical and experimental 
model for cosmological field theories. 

First we discuss the gauge invariance of the 
effective action in the presence of moving dislocations. 
It has been widely accepted that the condensate can be described by 
a hydrodynamic action, which is 
valid on a length scale larger than $\xi_{\alpha}$'s 
\cite{Hayashi-Yoshioka-u}, with $\xi_{\alpha}$ being
the amplitude coherence length
($\alpha = x,y,z$ depending on the direction). 
The action consists of the elastic part $S_{\rm el}$ and the 
interaction part with the electromagnetic field $S_{\rm int}$,  
%------------------------------------------------
\begin{eqnarray}
S &=& S_{\rm el} + S_{\rm int}\\
S_{\rm el} &=& \int d \tau d{\bf r} \frac{K}{2} 
\left\{
c_{0}^{-2}
\left(\partial_{\tau} \theta \right)^2 + 
\left(\partial_{x} \theta \right)^2 \right. 
\nonumber \\
&&\phantom{ \int d \tau d{\bf r} \frac{K}{2}\{\{}
\left.+ \gamma_{y}^{2} 
\left(\partial_{y} \theta \right)^2 + 
\gamma_{z}^{2} 
\left(\partial_{z} \theta \right)^2 \right\}\\
S_{\rm int} &=& \int d \tau d{\bf r}\ i L
\left( \varphi \partial_{x} \theta
+  A_{x} \partial_{\tau} \theta \right)
\label{action} 
\end{eqnarray}
%------------------------------------------------
where $\theta (\tau, {\bf r})$, $\varphi (\tau, {\bf r})$ 
and $A_{x}(\tau, {\bf r})$ 
are the phase of the order parameter, 
the scalar potential and the $x$-component of the 
vector potential, respectively. 
${\bf r}=(x,y,z)$ and $\tau$ is the imaginary time. 
The conducting chains are parallel to the $x$-axis. 
$K = N_{\bot} f_{\rm c}\hbar v_{\rm F}/ 2 \pi$ and 
$L = e N_{\bot} f_{\rm c}/\pi$,
where $f_{\rm c}$, $v_{\rm F}$, $N_{\bot}$ 
are the condensate fraction, 
the Fermi velocity parallel to the chains 
and the areal density of the chains, respectively. 
$c_{0} = v_{\rm F} \sqrt{m/m^*}$ is the phason velocity with 
$m^{*}/{m}$ being the effective mass ratio of the 
condensate to the normal electrons. 
$\gamma_{y} (= \xi_{y}/\xi_{x})$ and 
$\gamma_{z} (= \xi_{z}/\xi_{x})$ are 
the anisotropy parameters. 

According to the action, Eq. (\ref{action}), 
the charge density $\rho_{\rm c} (\tau, {\bf r})$ and current density 
${\bf j}_{\rm c} (\tau, {\bf r})$ of the condensate are expressed by,
%----------------------------------------------------
\begin{equation}
\rho_{\rm c} = L \partial_{x} \theta, 
\ \ 
{\bf j}_{\rm c} = - L\ {\hat{\bf e}}_{x}\ \partial_{\tau} \theta, 
\label{charge-current}
\end{equation}
%----------------------------------------------------
where ${\hat{\bf e}}_{x}$ is the unit vector parallel to $x$-axis. 
The charge conservation law then reads, 
%----------------------------------------------------
\begin{equation}
L \left(
\partial_{\tau} \partial_{x} \theta - 
\partial_{x} \partial_{\tau} \theta
\right)
= \Gamma_{\rm qp}. 
\label{charge-conservation}
\end{equation}
%----------------------------------------------------
Although $\Gamma_{\rm qp} (\tau, {\bf r})$ vanishes when 
$\theta (\tau, {\bf r})$ 
is a single valued smooth function of $\tau$ and $\bf r$, 
it does not vanish in the presence of 
moving dislocations \cite{Volovic} and 
gives \cite{core}
%----------------------------------------------------
\begin{equation}
\Gamma_{\rm qp} = - 2 \pi L \sum_{\nu} \int d l 
\left(\partial_{l} {\bf R}_{\nu} \times
\partial_{\tau} {\bf R}_{\nu}\right)_{x}
\delta^{(3)}({\bf r} - {\bf R}_{\nu}), 
\label{gamma}
\end{equation}
%----------------------------------------------------
where $\delta^{(d)}({\bf r})$ and ${\bf R}_{\nu}(\tau, l)$ are 
the $d$-dimensional $\delta$-function and 
the positions of the $\nu$-th dislocation line at 
imaginary time $\tau$, respectively. 
$l$ is a parameter along the line. 
In this Letter we consider only the dislocations with 
$2 \pi$ phase singularity and those with higher 
($4 \pi$, $6 \pi \cdots$) singularities are disregarded 
because of the higher excitation energies. 
The Eq. (\ref{gamma}) clearly shows that charge conservation 
is violated in the cores of the dislocations, and 
implies the conversion of 
the quasiparticles into condensate electrons. 
Although this phenomenon has been discussed earlier 
by several authors \cite{Ong-Maki,Brazovskii}, 
its effects on the effective action of the dislocations 
have not been considered. 

\begin{figure}
\begin{center}
{\parbox{6cm}
{\epsfysize=6cm \epsfbox{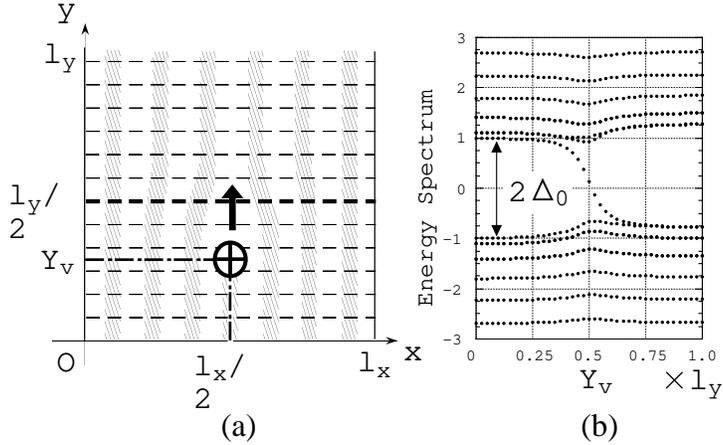}}}
\end{center}
\caption{(a) The configuration of the 
dislocation ($\oplus$), the chains (dotted lines) 
and wave fronts (shades). 
(b) The energy spectrum of a chain (bold line of (a)) 
as a function of the position of the dislocation, $Y_{v}$, 
is shown. 
All the energy levels are doubly degenerated at 
$Y_{v} = 0, l_{y}$ except for the innermost two with respect to the 
energy gap at $Y_{v}=0$. In this calculation $\xi = 0.1$ is taken. 
}
\label{spectral-flow}
\end{figure}

The origin of  $\Gamma_{\rm qp}$ 
becomes clearer by examining the energy spectrum of the quasiparticles 
in the core region. 
We consider chains of a finite length $l_{x}$ ($0 < x \leq l_{x} 
\equiv 1$) 
and assume that a straight dislocation line parallel 
to the $z$-axis is moving from $y = 0$ to $y = l_{y}$ 
along the plane $x = l_{x}/2$. (See Fig. \ref{spectral-flow} (a).) 
We calculate the energy spectrum of the chain located at 
$y = l_{y}/2$. 
The complex order parameter, $\Delta (\tau, {\bf r})$, on the chain 
is parameterized as $\Delta (x,Y_{v})$, 
where $Y_{v}$ is the $y$-coordinate of the dislocation line. 
The anisotropy of the system and the interchain electron hopping 
is neglected for simplicity. 
We take $\Delta (x,Y_{v})$ to be constant at $x = 0$, $1$, 
by assuming perfect pinning at the ends of the chains. 
$\Delta (x,Y_{v})$ must be quasiperiodic in $Y_{v}$, 
$\Delta (x,Y_{v}) \propto \exp (- 2 \pi i x)\ 
\Delta (x,Y_{v}+l_{y})$), corresponding to the addition of a 
wave front of CDW due to the dislocation \cite{Eilenberger}. 
By numerically diagonalizing the Hamiltonian, 
%-------------------------------------------------------
\begin{equation}
{\cal H} = \int d x\ \Psi^{\dagger}\cdot
\left(
\begin{array}{cc}
	-\hbar v_{\rm F}i \partial_{x} & \Delta (x,Y_{v})  \\
	\Delta^{*} (x,Y_{v}) & \hbar v_{\rm F}i \partial_{x}
\end{array} \right)\cdot\Psi, 
\label{Hamiltonian}
\end{equation}
%-------------------------------------------------------
where $\Psi = ({\cal R}(x),{\cal L}(x))$ with 
${\cal R}(x)$ and ${\cal L}(x)$ being 
the field operators for right and left moving electrons 
omitting the spin degree of freedom, 
we obtain the energy spectrum of the quasiparticles 
shown in Fig. \ref{spectral-flow} (b). 
It clearly shows the spectral flow phenomenon; one energy 
level is transferred from above to below the energy gap
through the core of the dislocation. 
When the energy level is occupied (vacant) at $Y_{v} = 0$, 
a quasielectron in the conduction band 
(quasihole in the valence band) is deleted 
(created) during this process. 
Since this process is strongly localized to the cores 
of the dislocations, the size of which 
is given by the amplitude coherence length $\xi_{\alpha}$'s, 
it cannot be described by the conventional 
hydrodynamic effective action of the condensate 
which is valid on a scale larger than $\xi_{\alpha}$'s. 

Our next task is to derive the effective action of the condensate 
which correctly describe the spectral flow in the cores of 
dislocations. 
We found that two correction terms 
to Eq. (\ref{action}) are needed. 
One is the correction term to the electromagnetic interaction, 
$S_{\rm int}$, 
and the other is related to the change of the total 
condensation energy by the spectral flow. 
The former should be introduced to compensate the 
violation of the gauge invariance discussed above. 
We first focus on a single chain to analyze 
this situation. 
The motion of a dislocation is then a phase 
slip process localized at the core \cite{Brazovskii}. 
The quasiparticles are created at this position and diffuse over the 
entire chain. 
The number of created quasiparticles is two (including the spin degree 
of freedom) at $0$ K, and decreases at higher temperatures in 
proportion to the equilibrium condensate fraction $f_{\rm c}$. 
The actual diffusion process of the quasiparticles is governed by the 
mobility, which depends on the material. 
In a dirty CDW sample, the motion of the quasiparticles is 
diffusive, whereas in a clean one it can be ballistic. 
The situations are classified by the relation between the 
quasiparticle velocity $v_{\rm qp}$ and 
the phason velocity $c_{0}$. 
The latter characterizes the velocity of 
the propagation of deformations 
in the condensate. 
The following two limiting cases are useful. 

{\it Clean limit} ($v_{\rm qp} \gg c_{0}$)\ --- \
Since the effective mass of 
the quasiparticles near the energy gap can be 
approximated by $m_{\rm eff} \simeq \Delta_{0}/v_{\rm F}^{2}$, 
where $\Delta_{0}$ is the equilibrium energy gap, 
an applied field accelerates the quasiparticles 
to $v_{\rm qp} = \sqrt{e \Delta_{0}/m_{\rm eff}} = v_{\rm F}$ 
without impurity scattering. 
(The particle-hole process limits the speed.) 
Since $v_{\rm F} \gg c_{0}$ the quasiparticles move much 
faster than the condensate in this limit. 
Then the spatial change in the quasiparticle current density 
at the dislocation position can compensate $\Gamma_{\rm qp}$ in 
Eq. (\ref{charge-conservation}) and 
the correction can be implemented by introducing the following 
term to the effective action, 
%----------------------------------------------------
\begin{equation}
\delta S_{\rm int} = - i L \int d\tau d{\bf r} \ 
A_{x} \left(\tau,{\bf r}\right)
\int^{x} dx'\ 
\Gamma_{\rm qp}\left(\tau,x',y,z\right). 
\label{clean-limit}
\end{equation}
%----------------------------------------------------

{\it Dirty limit} ($v_{\rm qp} \ll c_{0}$) \ --- \
With impurity scattering, $v_{\rm qp}$ is given by 
$- \mu_{\rm qp} E_{x}$, where $E_{x}$ and $\mu_{\rm qp}$ 
are the applied field
and quasiparticle mobility, respectively. 
% The mobility is related to the quasiparticle conductivity 
% $\sigma_{\rm qp}$ by $\mu_{\rm qp} = 
% \sigma_{\rm qp}/\rho_{\rm qp}$, 
% where $\rho_{\rm qp}$ is the 
% quasiparticle charge density \cite{Gruner}. 
When $v_{\rm qp}$ is much smaller than $c_{0}$ 
it is a good approximation to neglect 
the motion of the created quasiparticles. 
Then the correction term becomes,
%----------------------------------------------------
\begin{equation}
\delta S_{\rm int} = i L \int d\tau d{\bf r} \ 
\varphi \left(\tau,{\bf r}\right)
\int^{\tau} d\tau'\ 
\Gamma_{\rm qp}\left(\tau',{\bf r}\right). 
\label{dirty-limit}
\end{equation}
%----------------------------------------------------
For example, if the mobility is of the order of 
$\sim 1 \rm cm^{2}/{\rm V\cdot s}$, which is a typical value for 
the conventional samples, 
$v_{\rm qp}$ is smaller than $c_{0}$ unless $E_{x} > 10^{6} \rm V/cm$, 
which is much larger than the experimental value. 
The dirty limit therefore applies to most of conventional 
CDW samples. 

The second correction term arises from the condensation energy. 
The dislocation motion in CDW 
always produces quasiparticles from the condensate. 
If some quasiparticles already exist near the dislocation 
core, the dislocation can convert some to the condensate 
and gain condensation energy. 
Employing Eq. (\ref{gamma}), 
the contribution of the condensation energy 
to the effective action of the dislocation is estimated as 
$\delta S_{\rm cd} \sim \pm \int d \tau d {\bf r} (\Delta_{0}/e) 
\int^{\tau} d\tau' \Gamma_{\rm qp}(\tau',{\rm r})$, 
where $\pm$ depends on the situations described above. 
In most CDW materials, $\Delta_{0}/e$ is of the order of 
$0.1$ V and is important, especially when 
the $x$-component of the applied 
electric field is small (for example, in a transverse field 
condition \cite{Hayashi-Yoshioka}). 

Next we discuss the effective action of the dislocations taking 
into account the correction derived above. 
It is convenient to transform here to an 
isotropic coordinate system by 
$x = (x_{0},x_{1},x_{2},x_{3})$ with 
$x_{0} = c_{0}\tau$, $x_{1} = x$, $x_{2} = y/\gamma_{y}$, 
$x_{3} = z/\gamma_{z}$,
${\tilde K} = K \gamma_{y} \gamma_{z}/c_{0}$ and 
${\tilde L} = L \gamma_{y} \gamma_{z}/c_{0}$. 
We express $\partial_{\mu} \theta$ 
in terms of the dislocation coordinate since 
$\partial_{\mu} \theta$ is 
well-defined and single valued, although $\theta$ itself is not. 
As in Eq. (\ref{gamma}), we parameterize the position of the $\nu$-th 
dislocation in space and time by $l$ and $\tau$, respectively. 
The position of the line element in 4D space is expressed by 
$\eta_{\nu} (\tau, l) = (c_{0} \tau, {\bf R}_{\nu}) = 
(c_{0} \tau, X_{\nu}, Y_{\nu}, Z_{\nu})$. 
Introducing a flux tensor, ${\cal F}_{\alpha \beta}(x)$ 
\cite{Hayashi-Fukuyama}, 
%------------------------------------------------
\begin{equation}
{\cal F}_{\alpha \beta}(x) = 2 \pi \sum_{\nu}
\int d l d \tau \ 
(\dot{\eta}_{\nu})_{\alpha} (\eta'_{\nu})_{\beta}
\delta^{(4)} (x - \eta_{\nu}), 
\end{equation}
%------------------------------------------------
where $\dot{\eta}_{\nu} \equiv \partial_{\tau} \eta_{\nu}$ 
and 
${\eta'}_{\nu} \equiv \partial_{l} \eta_{\nu}$, 
we obtain 
%------------------------------------------------
\begin{equation}
\partial_{\mu} \theta (x) = 
\int d^{4} x' \epsilon_{\mu \alpha \beta \gamma}
\partial_{\alpha} G^{(4)} (x - x') 
{\cal F}_{\beta \gamma}(x'),
\label{deltheta}
\end{equation}
%------------------------------------------------
where $G^{(4)} (x) = -(4 \pi^{2})^{-1} |x|^{-2}$  
is the Green's function of 
$- \partial_{\mu}\partial_{\mu}$ and 
$\epsilon_{\mu \alpha \beta \gamma}$ is 
the fully antisymmetric tensor. 
Here the single valued part of $\theta$ is omitted. 
Employing Eq. (\ref{deltheta}) we can express all the terms 
in terms of ${\bf R}_{\nu}(\tau,l)$. 
For example, $S_{\rm el}$ of Eq. (\ref{action})
is rewritten as,
%------------------------------------------------
\begin{eqnarray}
S_{\rm el} &=& \int d^{4} x d^{4} x'\ \frac{\tilde K}{2}\
G^{(4)}(x - x') {\cal F}_{\alpha \beta}(x)
{\cal F}_{\alpha \beta}(x'), 
\label{dislocation-action-1}
\end{eqnarray}
%------------------------------------------------
where the conservation law of the topological charge of the 
dislocations, 
$\partial_{\alpha}{\cal F}_{\alpha \beta} = 0$
\cite{Hayashi-Fukuyama}, 
has been used. 

Now we consider the sliding of CDW by the nucleation of dislocations 
near the contacts \cite{Ramakurishuna,Duan,Maki}. 
The effective action discussed above depends on 
the electric field in the sample, 
which is conventionally assumed stationary and parallel to the chains. 
The \lq\lq two fluid approximation\rq\rq\ yields 
%------------------------------------------------
\begin{equation}
(-\partial_{x}^2 + \lambda_{0}^{-2}) E_{x} =
4 \pi D^{-1} \left\{{\bf j}_{\rm qp}\right\}_{x},
\label{electric-field}
\end{equation}
%------------------------------------------------
where $\lambda_{0}^{-2} = 8 e^{2} f_{\rm c}/(\hbar v_{\rm F})$, 
$D$ and $\left\{{\bf j}_{\rm qp}\right\}_{x}$, 
are the Thomas-Fermi screening length of the normal state, 
the diffusion constant and the $x$-component of 
the quasiparticle current, respectively. 
From this equation we can see that the electric field distribution 
depends on ${\bf j}_{\rm qp}$ and $D$, 
in other words, on the density and mobility of the quasiparticles. 
% In a metallic CDW such as $\rm Nb Se_{3}$, which has large 
% ${\bf j}_{\rm qp}$, $E_{x}$ is 
% almost constant throughout the sample, whereas in a semiconducting 
% CDW such as $\rm K_{0.3} Mo O_{3}$ and $\rm Ta S_{3}$, in which 
% ${\bf j}_{\rm qp} \simeq 0$, 
% $E_{x}$ is strongly screened in the bulk. 

The thermal and quantum nucleation rate of 
a dislocation loop in the presence of an applied electric field 
can now be estimated based on the effective action derived above. 
We employ the method by Langer and Fisher 
\cite{Langer-Fisher,Ramakurishuna,Maki}
and by Duan \cite{Duan}, to estimate the thermal and quantum 
nucleation rate, respectively. 
In the latter case 
we take the bounce solution of a dislocation 
loop as $\eta(\tau,\vartheta) = (c_{0} \tau , X, 
\sqrt{R^2 - (c_{0} \tau)^2} \cos \vartheta , 
\sqrt{R^2-(c_{0} \tau)^2} \sin \vartheta)$, 
($0 \leq \vartheta <2 \pi$), 
and use the instanton method. 

According to the above arguments we can distinguish
between the clean and dirty samples and 
between the metallic and semiconducting materials, 
which leads to the following picture.  

(1) In {\em clean systems}, in which the 
quasiparticle motion is ballistic, $\delta S_{\rm int}$ of 
Eq. (\ref{clean-limit}) must be employed. 
In this case the dislocations behave like an array of 
electric dipoles with moments aligned 
perpendicular to the chains \cite{Hayashi-Yoshioka} 
which do not couple with $E_{x}$. 
Therefore the nucleation of the dislocation loop does not lead to 
a gain of electrostatic energy. 
In addition, since the deviation of the quasiparticle density from 
the equilibrium is small, 
the gain of condensation energy is also negligible. 
The nucleation of the dislocation loops is then unlikely in these 
systems and sliding CDW transport is suppressed.

(2) In {\em dirty metallic systems}, 
$S_{\rm int}$ in Eq. (\ref{action}) is modified by 
Eq. (\ref{dirty-limit}) to
%------------------------------------------------
\begin{eqnarray}
S'_{\rm int} &=& S_{\rm int} + \delta S_{\rm int}\nonumber\\
&=&  i {\tilde L} \int d^{4} x\ \left\{
\varphi (x) \int^{x_{0}} d x'_{0}\ 
\partial_{1} \partial_{0}' 
\theta(x'_{0},{\bf x})\right.\nonumber\\
&&  \left.\phantom{  i L \int d^{4} x\ \ } + 
c_{0} A_{x}(x) \partial_{0} \theta (x)
\right\}\noindent\\
&\approx & \frac{i \pi {\tilde L}}{2} 
\Sigma_{\nu} \int d x_{0}
\int d l 
\left(
{\bf R}_{\nu} \times 
\partial_{l} {\bf R}_{\nu}
\right)_{x} V(X_{\nu}),
\label{gauge-invariant-action}
\end{eqnarray}
%------------------------------------------------
where $\partial_{1} \equiv \partial/\partial x_{1}$, 
$\partial_{0}' \equiv \partial/\partial x_{0}'$ and 
${\bf x} = (x_{1},x_{2},x_{3})$. 
In the last equation an approximation 
$c_{0} \rightarrow \infty$ is applied, 
which yields $G^{(4)}(x) \rightarrow 
-(4 \pi)^{-1} \delta(x_{0}) |{\bf x}|^{-1}$ and 
the gauge is fixed as $A_{x} = 0$. 
$V(X_{\nu})$ is given by $\int_{0}^{l_{x}} d x_{1}\ {\rm sgn}
(x_{1} - X_{\nu}) E_{x}(x_{1})$. 
Note from Eq. (\ref{gauge-invariant-action}) 
that $S'_{\rm int}$ is proportional 
to the area surrounded by the loop. 
In this case $E_{x}$ is approximately constant 
($\sim V/l_{x}$) in the entire sample, 
where $V$ is the applied voltage. 

Since the driving force of the dislocation is 
strongest near $x_{1} = 0$ and 
$x_{1} = l_{x}$, the nucleation of dislocation loop 
is most likely to occur at the edges. 
The effect of the boundary at $x_{1} = 0$ (or $l_{x}$) 
can be taken into account by considering mirror dislocations 
with respect to the boundaries, which replaces 
$V(X_{\nu})$ with $V^{b}(X_{\nu}) = 2 \int_{X_{\nu}}^{l_{x}} d x_{1}\ 
E_{x}(x_{1})$ ($= 2 \int_{0}^{X_{\nu}}\cdots$) 
in case of $X_{\nu} \sim 0$ ($\sim l_{x}$). 
The nucleation rate is then found to be 
proportional to $\exp \{-(V_{\rm Q}/V)^2\}$ 
for quantum nucleation, 
where $V_{\rm Q}^{2} = (\pi^{2}/12) \sqrt{m^{*}/m}\hbar^{3} 
\gamma_{y} \gamma_{z} v_{\rm F}^2 f_{\rm c} N_{\bot} e^{-2}$, 
and $\exp (-V_{\rm T}/V)$, for thermal 
nucleation, where 
$V_{\rm T} = \pi \gamma_{y} \gamma_{z} f_{\rm c} N_{\bot}
(\hbar v_{\rm F})^2/(e k_{\rm B}T)$. 
$V \approx |V^{b}(X_{\nu})|$ is the applied voltage. 

(3) In {\em dirty semiconducting systems}, 
the same action as Eq. (\ref{gauge-invariant-action}) applies. 
However, since ${\bf j}_{\rm qp} 
\sim 0$ in Eq. (\ref{electric-field}), 
$E_{x}$ in the bulk can be much smaller than $V/l_{x}$ 
due to the screening by the condensate, which reduces 
the nucleation rate. 
This qualitatively agrees with experiments 
\cite{ZaitzevZotov,Matsukawa}. 
Since the quasiparticle density is also strongly 
modulated, $\delta S_{\rm cd}$ becomes important, 
which causes a deviation of 
the nonlinear I-V characteristics from the metallic cases 
when $V < \Delta/e \simeq 0.1$ V. 

The result obtained in (2) coincides with 
those obtained earlier \cite{Ramakurishuna,Maki} 
in a different framework, 
which appears to be accidental however. 
These theories attribute 
the driving force of the dislocations 
to the elastic stress induced in 
the condensate by the applied electric field, 
which would imply the unlikely scenario that dislocations nucleate 
when an electric field is applied without contacts. 
From our theory it is clear that by nucleation of dislocations 
no energy is gained in the above situation, 
since it just results in an over-screening by the condensate. 
This demonstrates that our theory describes the 
screening and transport properties correctly. 

Before concluding this Letter, we comment on the similarity of the 
CDW system with field theories for the early universe. 
A dislocation in CDW is a counterpart of a cosmic string, 
{\it i.e.} a topological defect in electroweak field; 
The former creates quasiparticles 
from the condensate (Fermi sea) 
without creating anti-quasiparticles 
as is clear from Fig. \ref{spectral-flow}, 
whereas the latter produces matter without producing antimatter, 
which is called baryogenesis \cite{Dolgov,Diakonov}. 
Although the similarity of baryogenesis 
with the momentum creation in the 
vortices in superfluid He3 has already been discussed \cite{Bevan}, 
the CDW system is more similar to the electroweak theory 
in the sense that it has an identical 
chiral symmetry and chiral anomaly \cite{Su-Sakita}. 
We therefore expect that further 
theoretical and experimental studies of 
CDW may shed new light on its field theoretical counterpart
in particle physics and cosmology. 

In conclusion, we studied the dynamics of the dislocations 
in CDW's taking into account the contribution of the 
spectral flow in the cores. 
Our theory not only provides a natural basis to understand the 
dynamics of the dislocations in CDW's but also indicates 
new aspects of CDW. 

The author would like to thank 
G. E. W. Bauer for the critical reading 
of the manuscript and 
C. Dekker, Yu. V. Nazarov, 
H. S. J. van der Zant, B. Rejaei and M. I. Visscher 
for stimulating discussions. 
He is also grateful to H. Fukuyama and H. Matsukawa for useful 
comments. 
This work is part of the research program of the \lq\lq 
Stichting voor Fundamentele Onderzoek der Materie (FOM)\rq\rq\ 
which is financially supported by the \lq\lq Nederlandse Organisatie 
voor Wetenschappelijk Onderzoek (NWO)\rq\rq.

\vfill\eject

\end{document}